\documentclass[conference]{IEEEtran}
\IEEEoverridecommandlockouts
\usepackage{amsmath,amssymb,amsfonts}
\usepackage{algorithmic}
\usepackage{graphicx}
\usepackage{textcomp}
\usepackage[table]{xcolor}
\usepackage{booktabs}
\usepackage{tabularx}
\def\BibTeX{{\rm B\kern-.05em{\sc i\kern-.025em b}\kern-.08em
    T\kern-.1667em\lower.7ex\hbox{E}\kern-.125emX}}

\usepackage{xspace}
\newcommand{\ie}{i.e.,\xspace}
\newcommand{\eg}{e.g.,\xspace}

\newcommand{\etal}{et~al.\xspace}

\usepackage[autostyle, english=american]{csquotes}
\MakeOuterQuote{"}

\usepackage[hidelinks]{hyperref}
\usepackage{xurl}

\usepackage{cleveref}

\usepackage{enumitem}

\usepackage[most]{tcolorbox}
\newcommand*\circled[1]{\protect\tikz[baseline=(char.base)]{
            \protect\node[shape=circle,draw,inner sep=0.4pt,scale=1] (char) {#1};}}

\usepackage{lipsum}

 \usepackage[moderate,tracking=normal]{savetrees}

\begin{document}
\title{Designing Value-Based Platforms: Architectural Strategies Derived from the Digital Markets Act
\thanks{This research is partially funded by the Network Institute, Vrije Universiteit Amsterdam through the Network Institute Research Visit program.}
}

\author{\IEEEauthorblockN{Fabian Stiehle}
\IEEEauthorblockA{\textit{School of CIT} \\
\textit{Technical University of Munich}\\
Munich, Germany \\
fabian.stiehle@tum.de}
\and
\IEEEauthorblockN{Markus Funke}
\IEEEauthorblockA{
\textit{Software and Sustainability Group} \\
\textit{Vrije Universiteit Amsterdam} \\
The Netherlands
}
\and
\IEEEauthorblockN{Patricia Lago}
\IEEEauthorblockA{
\textit{Software and Sustainability Group} \\
\textit{Vrije Universiteit Amsterdam} \\
The Netherlands
}
\and
\IEEEauthorblockN{Ingo Weber}
\IEEEauthorblockA{\textit{School of CIT} \\
\textit{Technical University of Munich,}\\
\textit{Fraunhofer Gesellschaft}\\
Munich, Germany}
}

\author{	
\IEEEauthorblockN{
Fabian Stiehle\IEEEauthorrefmark{1},
Markus Funke\IEEEauthorrefmark{2},
Patricia Lago\IEEEauthorrefmark{2},
Ingo Weber\IEEEauthorrefmark{1}\IEEEauthorrefmark{3}
}
\IEEEauthorblockA{\IEEEauthorrefmark{1} School of CIT, Technical University of Munich, Germany, Email: fabian.stiehle@tum.de
}
\IEEEauthorblockA{
	\IEEEauthorrefmark{2} Software and Sustainability Group, Vrije Universiteit Amsterdam, The Netherlands
}
\IEEEauthorblockA{
	\IEEEauthorrefmark{3} Fraunhofer Society, Headquarters, Munich, Germany
}
}

\maketitle

\begin{abstract}
The digital markets act (DMA) regulates very large digital platforms like Meta's Facebook or Apple's iOS with the goal to promote fairness, contestability (of market power) and user choice. From a system design or broader technical perspective, the implications of the DMA have not been studied so far.
Using systematic methods from qualitative coding and thematic analysis, we investigate the DMA from a technical perspective and derive eight high-level design strategies that serve as fundamental approaches towards value-based architectural goals like 'fair practice', or 'user choice' (as envisioned by the DMA).
We investigate how compliance with the DMA has been achieved and derive 15 tactics that we map to our strategies. 
While the DMA obligations challenge existing platform designs, they also create new opportunities for designing services within these huge ecosystems. We, thus, discuss our strategies in light of both.
We see this work as a first step towards filling this pressing gap
in the architecture of platform ecosystems, i.e., how to incorporate abstract human values in architecture design.
\end{abstract}


\section{Introduction}
The platform economy has formed novel digital markets 
by connecting large numbers of users directly with businesses. Platforms form two-sided markets, where the attractiveness of one side (e.g., number of users) governs the attractiveness of the other side (e.g., to advertisers)~\cite{rochet2003platform}. Companies like Alphabet, Apple, or Meta are in command of very large platforms that form quasi-monopolies, controlling market access and leveraging strong lock-in effects~\cite{andersson2017platform}. 
Awareness of the downsides of such centralised (orchestration) power is rising, driven by large scandals like \textit{Cambridge Analytica}.
Platform businesses are criticised to harm fundamental societal values. Among other, for anti-competitive (unfair) behaviour~\cite{dolata2017apple}, for exploiting large amounts of personal data~\cite{zuboff2023age}, or even for contributing to the erosion of democratic norms and values~\cite{weinhardtDigitalDemocracyWakeUp2024}.

To tackle these issues from a policy side, regulatory bodies, such as the European Union, have enacted a range of new regulations like the \textit{General Data Protection Regulation} (GDPR)~\cite{gdpr}, or more recently the \textit{Digital Markets Act} (DMA)~\cite{dma}. 
The DMA targets the orchestration power of businesses in command of very large digital platforms, which it designates as so-called \textit{gatekeepers}. Gatekeepers are able to make use of very strong network and lock-in effects, as well as advantages driven by large amounts of collected data~\cite[Recital 2]{dma}. As such, these gatekeepers are impossible to contest on the merits of their products alone~\cite[Recital 5]{dma}. An example where gatekeepers leverage these advantages is \textit{vertical integration}. Where gatekeepers are in control not just of a single platform, but also of the services within these platforms. As they set the rules on their own platforms, they can favour their own offerings (e.g., by limiting access to certain core features to their own offerings or prominently positioning their services over competitors)~\cite[Recital 51]{dma}. Such behaviour is considered to be unfair by the DMA and to have detrimental effect on innovativeness, quality, and user choice~\cite[Recital 4]{dma}. Additionally, the DMA is based on the notion that gatekeepers have not continued to innovate, and are reaping more than their fair share (c.f.~\cite{bostoen2023understanding}), due to the low bargaining power of their competitors and platform users~\cite[Recital 4]{dma}.
Consequently, the DMA sets out to achieve values such as contestability (of market power), fairness, user choice, and innovation, by regulating (restricting or obligating) practices of gatekeepers \cite[Art 5-7]{dma}. 

Meanwhile, in research there has been a renewed call to integrate (human) values (like fairness) into the the engineering of information systems (e.g., \cite{pelliccione2023architecting,AlidoostiEtAl_ExploringEthical_2025, spiekermann2022value, van2007ict}). 
Such values need to be considered early in the design process, when `value considerations can still make a difference'~\cite[p. 4]{van2015design}. 
Consequently, there are calls to integrate value perspectives along with ethical reasoning within software architecture practices~\cite{AlidoostiEtAl_IncorporatingEthical_2022,wohlrab2024supporting}. This could address some of the aforementioned issues by uncovering (unintended) bad consequences early in the system design. However, how to translate abstract values into technical design concerns remains challenging~\cite{AlidoostiEtAl_ExploringEthical_2025}. 

In our work, we extract values embedded in the DMA and the associated design considerations (as obligated by the DMA) and systematise them into actionable \textit{design strategies}. 
Our work draws on the insights of Hoepman~\cite{hoepman2014privacy}, who translated the abstract notion of privacy into architecture design strategies~\cite{hoepman2014privacy}.
Such strategies describe a fundamental approach towards an architectural goal to achieve a certain aspect of a quality~\cite{hoepman2014privacy}. Hoepman’s privacy design strategies are an instantiation of such for the goal of \textit{privacy-by-design} to achieve a certain aspect of privacy protection. Colesky et al.~\cite{colesky2016critical} subdivided these strategies further into tactics in order to interpret the GDPR.
Thus, such strategies can be understood as ‘category of tactics', as in Bass et al. (c.f., for example their categories for ‘detect attacks', or ‘resist attacks’, to achieve the \textit{security} quality~\cite[Chp 11.2]{bass2021software}.) or as ‘design concerns’, as shown in Harrison and Avgeriou~\cite{harrison2010architecture}. To avoid confusion, we will adopt Hopeman’s terminology of \textit{strategies}. 

The DMA fits our method to derive values and strategies for four reasons: 
(i) the DMA is ‘preventative, not reactive’, it regulates unfair practices, rather than reacting to them~\cite{simone2025principles};
(ii) it imposes a concrete list of do's and dont’s with direct demands on the technical design~\cite{bostoen2023understanding}; 
and (iv) all gatekeepers are required to submit detailed compliance reports, outlining all measures they have implemented in response to each obligation.

We use systematic methods of qualitative coding~\cite{saldana2021coding} to derive values, problems, and design considerations (i.e., do’s and dont’s). We systematise these into eight strategies and match these against the obligations imposed by the DMA. We then investigate how gatekeepers implement these obligations by deriving \textit{gatekeeper tactics}. We investigate reports of Alphabet, Amazon, Apple, Booking, and Meta, to cover all major platform types, comprising around 650 pages of compliance documentation. We match our derived tactics to our strategies.

These strategies and tactics can guide future (re-)designs of platforms to achieve values like fairness and user choice, as envisoned by the DMA.
For example, to design for a value like \textit{`fair practice'}, a strategy \textit{`Equal Access \& Interoperability'} could be employed, which we derived from the DMA, and demands that all services competing on the platform have the same access to the platform features, its users, etc. This has consequences for the whole architecture. One of the offered tactics by the gatekeepers is \textit{`Alternative App Distribution'}, which allows apps to be installed on the platform through alternative stores and websites. This concerns the architecture, as it requires to design for alternative deployment methods and full integration with system components, while maintaining security.

While the gatekeeper tactics correspond to obligations on the gatekeepers side, they also correspond to opportunities and affordances for all other services and apps operating in a gatekeepers ecosystem (i.e., which new design options emerge for third-parties based on a chosen tactic). Therefore, tactics for platform design must incorporate the inherent two-sidedness of a platform. For example, the aforementioned tactic \textit{`Alternative App Distribution'}, may allow to distribute their app via their existing deployment and distribution channels, instead of going through the gatekeeper's systems.
We thus also discuss the affordances gatekeeper tactics create.

While software architecture has shown that it can effectively integrate values with ethical import—such as privacy~\cite{colesky2016critical} or sustainability~\cite{lago2025sustainability}—the negative societal impacts of software platforms have so far been neglected. We see this work as a first step towards filling a pressing gap in the integration of human values in the architecture of platform ecosystems.
\section{Background and Related Work}
\label{sec:rw}
\subsection{Digital Platforms}
Platforms are studied from a wide variety of perspectives~\cite{hein2020digital,de2018digital}. Most conceptualisations reference the seminal paper of Rochet and Tirole~\cite{rochet2003platform}, concerning the two-sidedness of platform markets,
i.e., how the size of one side of the market (e.g., amount of end-users) has a direct influence on the attractiveness of the other side (e.g., for advertisers). 
Software engineering (SE) studies platforms in the context of \textit{software platforms} embedded in \textit{software ecosystems}~\cite{jansen2009sense}. 
In literature, many definitions for software ecosystems are adopted
-- for an overview, see Manikas and Hansen~\cite{manikas2013software}. 
Within SE, research has focused on technical issues like modularity, evolution, or integration~\cite{manikas2016revisiting}.
In software architecture, work exists on the requirements of industry platforms (e.g.,~\cite{eklund2013characterising,de2016study}), evaluation of ecosystem architectures~\cite{anvaari2010evaluating,amorim2015tailoring}, or the general challenges of architecting for ecosystems~\cite{bosch2010architecture}, among other.
The possible societal negative impacts have been neglected. Consequently, a growing list of works are calling for research on the integration of human values into the architecture of ecosystems~\cite{pelliccione2023architecting,gordijn2024blockchain}. However, this has not yet materialised.
Closest to our work are works studying the openness~\cite{jansen2013quality,anvaari2010evaluating} or transparency of ecosystems~\cite{cataldo2010architecting}. However, openness is limited to an ecosystem developer's point of view, while transparency is limited to technical interfaces.

\subsection{Software Design for Human Values}
\noindent Value Sensitive Design (VSD)~\cite{friedman2013value} can be considered the most mature approach to integrate human values in software engineering (c.f., Alidoosti et al.~\cite{AlidoostiEtAl_ExploringEthical_2025}) 
and has given rise to an \textit{IEEE} standard for value-based engineering~\cite{spiekermann2022value}. This standard presents a method to derive so-called ethical value requirements. To move from  problem to solution space, Wohlrab et al. have argued for value tactics, which are concrete design decisions to promote a certain value~\cite{wohlrab2024supporting}. How to move from abstract values to concrete requirements remains challenging. A recent systematic review found only two studies considering this aspect~\cite{AlidoostiEtAl_ExploringEthical_2025}.
For software architecture specifically, Alidoosti et al.~\cite{AlidoostiEtAl_IncorporatingEthical_2022} calls for the integration of ethical reasoning during software architecture and finds current methodologies lacking. 
Closest to our work is the work by Hoepman~\cite{hoepman2014privacy}, which translates the abstract notion of privacy into concrete design strategies. 
Design strategies describe fundamental approaches to the design of a system to achieve a certain goal. They can be compared to categories of tactics, as in Bass et al.~\cite{bass2021software}.
Hoepman's privacy design strategies are an instantiation of strategies for the goal of privacy-by-design~\cite{hoepman2014privacy}. Colesky et al.~\cite{colesky2016critical} subdivide these strategies further into tactics to interpret the GDPR, which obligates privacy by design.
\begin{figure*}[t]
    \centering
    \includegraphics[width=.8\linewidth]{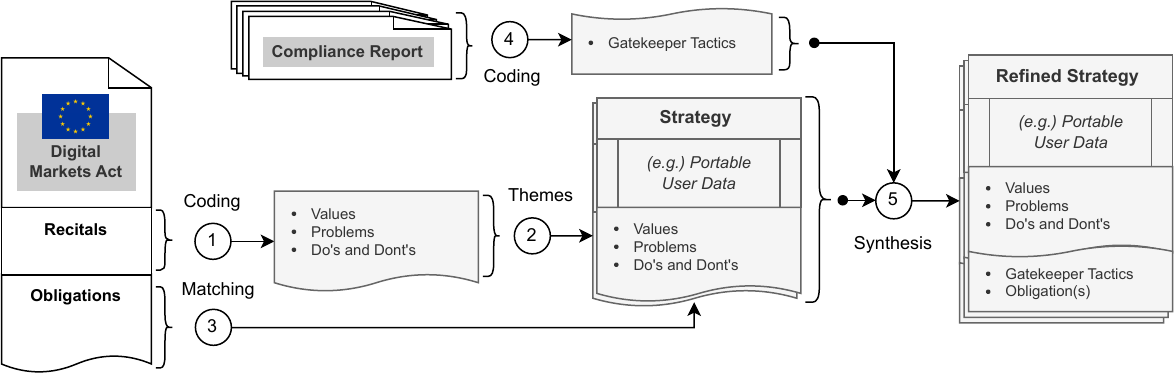}
    \caption{Study design overview, with phases \circled{1} - \circled{5}}
    \label{fig:studyDesign}
\end{figure*}
\subsection{Digital Markets Act}
\label{sec:theDMA}
The digital markets act (DMA)~\cite{dma} was adopted in September 2022 as \textit{Regulation on contestable and fair markets in the digital sector}. The DMA is part of a suit of regulations targeting digital transformation. It is closely related to the \textit{Digital Services Act}, which targets accountability of online platforms regarding their content. The DMA is often analysed from the standpoint of competition law. Indeed, most of its obligations can be traced back to previous antitrust cases across the EU~\cite{bostoen2023understanding}.
However, the DMA differs from antitrust regulations in the sense that it is preventative, not reactive; i.e., it regulates unfair practices, rather than reacting to them~\cite{simone2025principles}. It defines a relatively precise list of do's and dont’s (\cite[Art. 5, 6, 7]{dma}, c.f.,~\cite{bostoen2023understanding}) regarding the practices of so-called \textit{gatekeepers} on their \textit{core-platform services} (e.g., search engines, operating systems, cloud computing services, or application stores)~\cite[Art 2(1)]{dma}. Gatekeepers are meant to be (i) an important gateway for business users to reach end users, (ii) have significant market impact, and (iii) have an 'entrenched and durable position'.
These criteria are primarily assessed by some quantitative thresholds~\cite[Art 3]{dma}. 
The current list of gatekeepers includes companies like \textit{Alphabet} (e.g., Search, Android), \textit{Apple} (e.g., App Store, iOS), \textit{Booking}, or \textit{Meta} (e.g., Facebook, Whatsapp).~\cite{EuropeanCommission2025Gatekeepers}.
The DMA, thus, focuses on businesses whose platforms have such significant impact on the market that it warrants regulation~\cite[Recital 15]{dma}. However, it is stressed that smaller platform businesses may employ the same prohibited practices to reach an entrenched position eventually, in which case an intervention may be warranted~\cite[Recital 25]{dma}. The DMA further calls for compliance by design. Specifically, as it has to be integrated with other regulations like the GDPR~\cite[Recital 65]{dma}.

The DMA's obligations revolve around the values of contestability and fairness (c.f., also~\cite{bostoen2023understanding,simone2025principles}). \textit{Contestability} refers to the fact that any player on a market must remain contestable based on the merits of their product~\cite[Recital 32]{dma}; while \textit{fairness}, in the context of the DMA, is concerned with the balance between rights and obligations of a platform compared to its business users (called \textit{fair practice} thereafter)~\cite[Recital 33]{dma}. These values can also be seen in that they address concerns of inter (between platforms) and intra-platform (the platform's internal market) competition. Specifically, the DMA sees a causal connection between them. Lack of contestability can lead to a gatekeeper employing unfair practices on their market. This, in turn, further strengthens the uncontested position of the gatekeeper~\cite[Recital 34]{dma}. This comes to a detriment of overall fair market outcomes (in terms of prices, quality and choice)~\cite[Recital 107]{dma}. Other values we extracted relate to, e.g., innovation, user choice, multi-homing, or transparency.
While already enacted in 2022, the effects of the DMA are still unclear. There are still ongoing non-compliance investigations, and an ongoing debate in literature about its effectiveness or possible blind spots (see e.g.,~\cite{cennamo2022value,simone2025principles}).

Gatekeepers are obligated to publish yearly compliance reports, where they have to outline in detail their measures to implement the obligations~\cite{EuropeanCommission2025CompRep}. 
While we take a value perspective, and focus on abstract design strategies~\cite{hoepman2014privacy} for general platform design, compliance by design methods~\cite{colesky2019helping} can be seen as orthogonal to our work. However, the DMA has not yet been considered from such, or any, technical perspective. Closest to our work is a recent study outlining possible designs of interoperable messaging solutions~\cite{len2023interoperability}, as mandated by Art 7 of the DMA. 
\begin{figure*}[ht]
    \centering
    \includegraphics[width=0.85\linewidth]{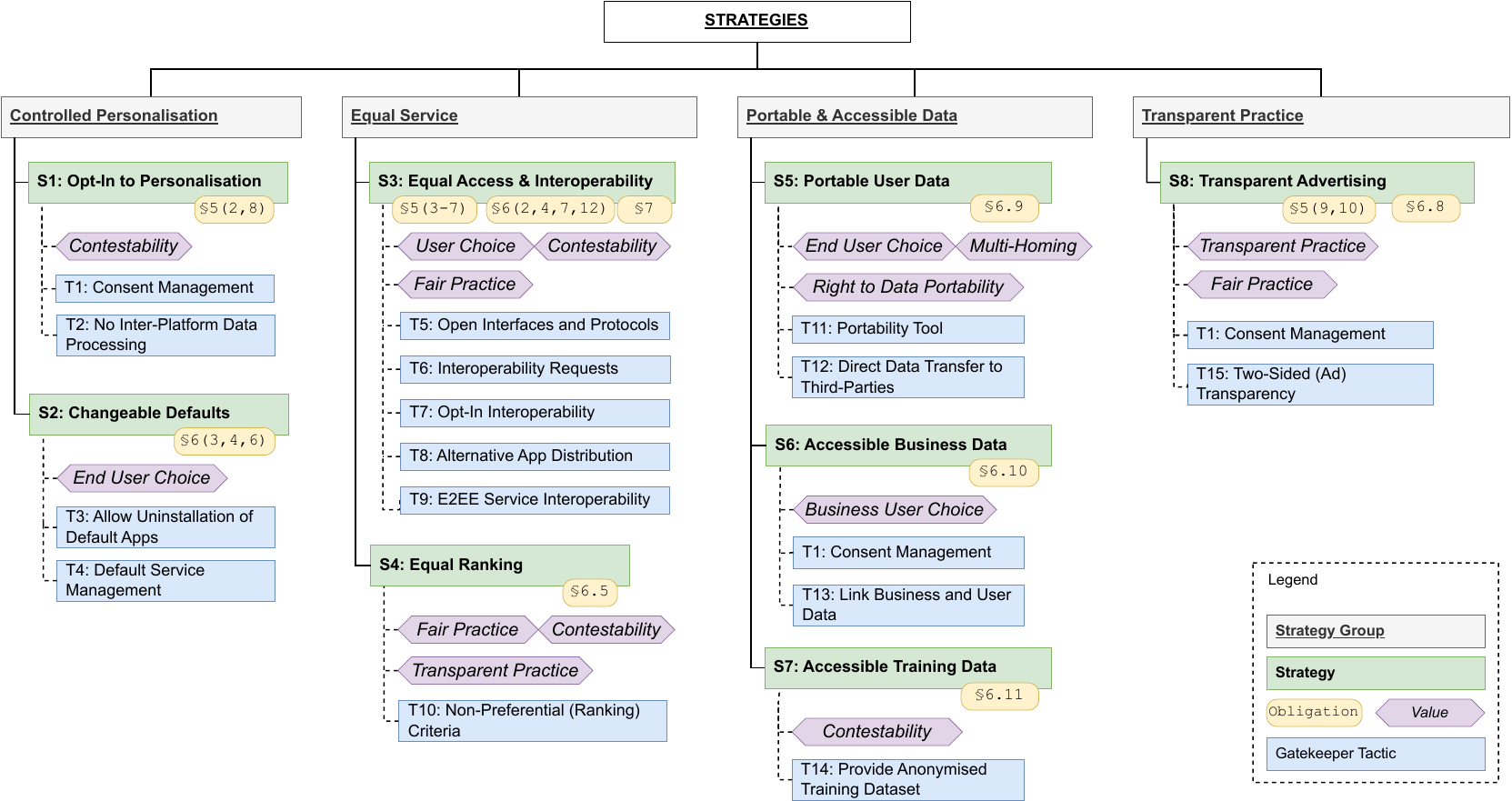}
    \caption{Overview of extracted strategies (green boxes), related obligations (rounded yellow boxes), values (diamond boxes), and tactics (blue boxes).}
    \label{fig:octopus}
\end{figure*}
\section{Methodology}
\label{sec:method}
%

In our study, we follow the recommendations for qualitative coding by Saldaña~\cite{Saldana_CodingManual_2016} and for thematic analysis in software engineering by Cruzes and Dyba~\cite{CruzesDyba_RecommendedSteps_2011}. Our study process is illustrated in \Cref{fig:studyDesign} and comprises four distinct phases, which are explained below. 
In summary, \circled{1} we coded the DMA recitals to extract problems, intentions, and values, \circled{2} we performed a thematic analysis to arrive at a set of preliminary strategies, \circled{3} we matched the DMA's obligations to our strategies to assess completeness, 
\circled{4} we derived tactics from available gatekeeper compliance reports, which we also matched to our strategies, and finally \circled{5} we composed our results to our final strategies. 
Potential threats to validity are discussed in \Cref{sec:ttv}.
To follow standards of open science and to increase reproducibility, we provide a replication package, including our study artefacts and intermediate results (\eg complete coding book and final strategy collection).\footnote{Our  replication package can be accessed at \url{https://github.com/fstiehle/rp-dma-architecture-strategies}. An archived version is available at \url{https://doi.org/10.5281/zenodo.18620024}.}


\subsection*{Phase \circled{1} - Coding of Recitals}
The DMA's 109 recitals interpret and contextualise the obligations (e.g., spelling out values, and problems the obligations are trying to address). We apply coding techniques to extract: (i) the \textbf{problems} the DMA aims to solve; the (ii) \textbf{intent} behind the rules, clarifying how obligations should be interpreted; and (iii) the underlying \textbf{values}, which the obligations are concerned with. Two distinct coding cycles were applied.

In the first cycle, for each recital, we applied sub-coding to identify problems, intent and values~\cite{Saldana_CodingManual_2016}. To derive problems we applied open-coding, for example \textit{'increases costs for end users to switch'}. To determine the \textbf{intent}, we distinguished between \textit{do's} and \textit{dont's}. The \textit{do's} are actionable, typically marked by keywords such as \textit{'enable'} or \textit{'ensure'}, while \textit{dont's} are constraining, with keywords like \textit{'should not restrict'}. To derive \textbf{values}, we primarily applied \textit{in vivo} coding, to extract values in an unfiltered manner, such as \textit{'fair market outcomes'} or \textit{'contestability'}.
We concluded this phase with a second cycle, applying axial coding~\cite{Saldana_CodingManual_2016} to uncover categories of codes (e.g., similar problems, different mentions of fairness).



\subsection*{Phase \circled{2} - Thematic Analysis}
Based on our derived codes, we now applied a thematic synthesis~\cite{CruzesDyba_RecommendedSteps_2011} to form preliminary strategies. 
Intents (\ie do's and dont's) were synthesised into higher-level themes that constitute our actual \textbf{strategies}. For instance, the strategy \textit{'Changeable Defaults'} captures do's such as \textit{'Allow uninstall of default applications.'} and dont's such as \textit{`Don't Prevent the Switch'}. 
The strategy becomes part of the broader theme \textit{`Controlled Personalisation'}

\subsection*{Phase \circled{3} - Matching Obligations to Strategies}
As recitals give context to all obligations, no new strategies should emerge from the obligations alone. We used this as an additional validation step by matching all obligations imposed on the gatekeepers (Art 5-7) to our strategies. In case an obligation could not be satisfied by our strategies, that would indicate that we might need to revise or extend them. This case did not occur.  

\subsection*{Phase \circled{4} \& \circled{5} - Compliance Report Coding \& Synthesis}
Proposing an ideal tactic at this stage is not feasible due to the lack of empirical research on tactics satisfying the DMA in the best possible way. Instead, we derived tactics from gatekeeper's publicly accessible compliance reports. We call these \textit{gatekeeper tactics}. 
Using the gatekeepers' compliance reports provides us with three benefits: (i) it introduces a second validation step, allowing us to verify whether our strategies cover all tactics (gatekeepers have to describe their response to all obligations), (ii) it offers an initial set of tactics derived directly from the gatekeepers' implementations, opening avenues for future research for evaluation, and (iii) it allows us to uncover the affordances these tactics create for systems and applications within these ecosystems. Not all measures outlined in the reports have technical relevance; some focus only on contractual terms and were therefore not considered for extraction. For example, Art 6(13) requires gatekeepers to not have disproportionate contractual termination conditions with their users.


We considered compliance reports from Alphabet, Amazon, Apple, Booking, and Meta, covering all platform categories and comprising around 650 pages of compliance documentation.
We applied two coding cycles on the gatekeeper reports. First, we used deductive coding~\cite{CruzesDyba_RecommendedSteps_2011} to match our existing strategies. 
During the same cycle, we applied open coding~\cite{Saldana_CodingManual_2016} to derive an initial set of tactic names and short summaries.

In the second cycle, we applied axial coding 
to identify common tactics.
We then documented the resulting tactics.
This phase resulted in a set of 15 gatekeeper tactics, each documented by the following elements: name, description, architecture influence, and affordance perspective.

In the final step \circled{5} of our study, we synthesised the intermediate results, \ie the preliminary strategies and gatekeeper tactics, into consolidated strategies. An overview of our result is given in Figure~\ref{fig:octopus}. This phase concludes our research with a final collection of eight strategies and a total set of 15 gatekeeper tactics.

%
%
\section{Results}
Figure \ref{fig:octopus} gives an overview of our extracted strategies, their related obligations, values, and tactics. 
In Table \ref{tab:tactics}, we show a summary of extracted gatekeeper tactics with their description, architecture influence, and affordances they create. In this short paper, we will focus on outlining the overarching strategies. Following, we will discuss each strategy in more detail in terms of the addressed problems (coded from the DMA), and considerations (do's and dont's).

For the remainder of the paper, we loosely denote a software-intensive system, which is operated by a business user within the platform ecosystem, a \textit{service}. An \textit{app} is a special kind of service which can be installed on a platform (usually an operating system platform) and is operated by a business user. A \textit{platform service} is a vertically integrated service, i.e., it is operated by the same entity as the platform. Finally, a \textit{platform} is a service that directly connects end users with business users.

\renewcommand{\arraystretch}{1.2}
\begin{table*}[t]
\centering
\footnotesize
\caption{Overview of Derived Gatekeeper Tactics}
\label{tab:tactics}

\rowcolors{2}{white}{gray!10}
\begin{tabularx}{\linewidth}{p{160pt} p{180pt} X}
\toprule
\textbf{Tactic Name and Description} & \textbf{Architecture Influence} & \textbf{Affordance Perspective} \\
\midrule

\textbf{T1: Consent Management} \newline
Allow Users the choice on how their data is processed. & 
Consent screens, consent management and consent enforcement for each user must be incorporated throughout the whole system architecture. & 
-- \\

\textbf{T2: No Inter-Platform Data Processing} \newline 
Do not cross-use data of users in the first place. & 
All data flows need to be inspected and restricted to a local context, i.e., they cannot be cross-used. & 
-- \\

\textbf{T3: Allow Uninstallation of Default Apps} \newline 
Allow users to uninstall a platforms default apps, when not system relevant. & 
Requires a clear separation of critical and non-critical components. Requires increased modularity. & 
-- \\

\textbf{T4: Default Service Management} \newline 
Allow different default services for categories of functionalities. & 
Requires separation and modularity of components. Requires user choice screens, system wide default registries, and additional APIs to support third-party default services. &
Prompt a user to become its new default app. \\

\textbf{T5: Open Interfaces and Protocols} \newline
All platform features are made equally accessible to its services. & 
Features must be designed with open interfaces and protocols in mind. & 
Gain equal access to platform features. \\

\textbf{T6: Interoperability Requests} \newline
Critical interoperability features are accessible by request and have attached requirements. & 
Interoperability features need to be encapsulated and require an access management, which lifecycle (requesting, granting, using, and revoking) need to be managed. & 
Gain access to additional platform features (e.g., default application store, hardware features). \\

\textbf{T7: Opt-In Interoperability} \newline 
Interoperability with core services (e.g., identity, payment processing, distribution) is optional when feasible, with support for partial integration. & 
Architecture requires increased modularity, separation of non-critical and essential components, and interfaces that support partial integration. & 
Use alternatives to core services for, e.g., payment, sing-on, or analytics. \\

\textbf{T8: Alternative App Distribution} \newline 
Allow apps to be installed through alternative stores and websites. & 
Requires alternative deployment methods of apps and full integration with system components, while maintaining security. & 
Offer apps through alternative distribution methods (e.g., through your website). \\

\textbf{T9: E2EE Service Interoperability} \newline 
Messenger services integrate with third-party services, while still ensuring E2E-Encryption. & 
Requires the integration of interoperable security protocols and architectures. & 
Interoperate with an existing messenger service to reach a larger user-base.  \\

\textbf{T10: Non-Preferential (Ranking) Criteria} \newline 
Apply equal and transparent criteria to any algorithms affecting the visibility or placement of advertising, products, or services. & 
Features related to indexing, crawling, rating, or search must be designed with non-preferentiality in mind. Results must be explainable and verifiable. & 
Equal placement and exposure on the platform. \\

\textbf{T11: Portability Tool} \newline 
Allow users to (selectively) export their data via a tool or dashboard. & 
Requires a central portability component and all user data to be accessible for export in machine readable format. & 
Import existing user data from a platform to, e.g., reduce switching cost for the user. \\

\textbf{T12: Direct Data Transfer to Third-Parties} \newline 
Allow users to (selectively) directly transfer their data to an authorised and registered third-party. & 
Requires an API to handle authentication, authorisation, and data transfer. & 
Automatically and continuously import existing user data from a platform to, i.e., allow multi-homing. \\

\textbf{T13: Link Business and User Data} \newline 
Allow business users to access their data and the data of their customers. & 
Requires accessibility features (e.g., dashboards, APIs, exports) and all business user data and directly related customer data to be identified, linked, and be made accessible. & 
Gain access and insights into customer data. \\

\textbf{T14: Provide Anonymised Training Dataset} \newline 
Provide access of data collected on the platform used to train search and ranking algorithms, without compromising the privacy of users. &
Data collected on the platform (e.g., on query, click, view, and ranking data) used to train algorithms needs to be anonymised and made accessible to third-parties. &
Request access to training data to improve algorithms. \\

\textbf{T15: Two-Sided (Ad) Transparency} \newline 
Allow independent verification of remunerations received through advertising. & 
Requires collection and accessibility (through dashboards, APIs, exports) of all relevant ad data across all services for both publishers and advertisers. & 
Transparency on ad performance, gatekeeper fees, and any possible deductions and surcharges. \\

\bottomrule
\end{tabularx}
\end{table*}

%
\subsection{Controlled Personalisation}
This strategy group emphasises user control on the level of personalisation  to increase contestability, by (i) making use of  personalised data optional, and (ii) giving users the necessary tools to personalise their experience.
\subsubsection{\textbf{S1: Opt-In to Personalisation}}
\label{sec:optInStrat}
Increased personalisation (e.g., through cross-using data between services) is optional, a user can choose a less personalised alternative. \textit{Problems:} Gatekeeprs can gain an unfair advantage through the accumulation of data across multiple platforms or services; if this practice is obligatory, it creates lock-in effects.
\textit{Considerations:} The DMA requires gatekeepers to only cross-use data with user consent. Further, if no consent is given, an 'equivalent alternative' must be provided~\cite[Art 1(2)]{dma}. Dont's prohibit practices that force users to signup with a platform as condition of accessing a particular service~\cite[Art 1(8)]{dma}. 
\subsubsection{\textbf{S2: Changeable Defaults}} The system provides the necessary means to allow a user to change its defaults.
\textit{Problem:} Gatkeepers can use their position to favour certain (e.g., its own) services to the detriment of competitors.
\textit{Considerations:} The DMA requires that pre-installed (non-critical) software can be uninstalled. Further, that any default settings which favour own services can be changed~\cite[Art 6(3)]{dma}. Specifically, the DMA asks for choice screens for search engines, virtual assistants, or web browsers. Dont's prohibit conditions that would prevent users from switching (or offering) competing services~\cite[Art 6(3)]{dma}. 
This strategy has synergies with \textit{S3: Equal Access \& Interoperability}. As third-party default services (e.g., an app store) will require interoperability with platform features. Thus, some tactics overlap.
%
\subsection{Equal Services}
This group of strategies emphasises fairness, by promoting practices that ensure that all services on the platform have (i) equal access to platform markets and features, (ii) and are equally ranked on the platform.
\subsubsection{\textbf{S3: Equal Access \& Interoperability}}
\label{sec:equalInterop}
All services competing on the platform have the same access to the market, its users, and features.
\textit{Problems:} Platforms are often vertically integrated, where a gatekeeper takes on a dual-role. If this position is used to give certain (e.g., its own) services preferred or unique access to operating, hardware, or software features, it will undermine innovation and user choice. Furthermore, the lack of interoperability can lead to strong network effects, as e.g., for messenger services, reducing contestability and increasing the cost for users to switch.
\textit{Considerations:} The DMA requires that platform features are equally accessible to all services~\cite[Art 6(7)]{dma}. Further, it requires support for alternative app distribution methods~\cite[Art 6(4)]{dma} and interoperability of messenger apps~\cite[Art 7]{dma}. Finally, equal interoperability goes both ways in the way that a service cannot be forced to interoperate with a platform service~\cite[Art 5(7)]{dma}.
Equal accessibility also comes with a range of dont's which do not have direct technical relevance, focusing on practices which can limit alternative choice~\cite[Art 5(3-5)]{dma}, or which would use data, only available to the platform, to gain a competitive advantage for a platform service~\cite[Art 6(2)]{dma}. Other dont's target contractual terms regarding the fairness of access~\cite[Art 6(12)]{dma}, termination conditions~\cite[Art 6(13)]{dma}, and the seeking of redress~\cite[Art 5(6)]{dma}.

\subsubsection{\textbf{S4: Equal Ranking}}
Visibility on the platform is non-discriminatory, and all services (and their content) are treated equally in placements like listings or search results.
\textit{Problems:} Gatekeepers can use their position to favour their own services, apps, or content over that of third parties, when it comes to listings, placement, and general visibility.
\textit{Considerations:} 
The DMA obligates ranking to be `fair' and 'transparent'~\cite[Art 6(5)]{dma}, where with ranking 'all forms of relative prominence, including display, rating, linking or voice results' are considered~\cite[Recital 52]{dma}. 

\subsection{Portable \& Accessible Data}
Access to data constitutes a significant advantage and should not be locked in. This strategy emphasises accessibility of (i) end-user, (ii) business-user, and (iii) training data.
\subsubsection{\textbf{S5: Portable User Data}} User data is made exportable and transferable to other services.
\textit{Problems:}
Platforms are in a unique position to collect and benefit from large amounts of user data. Restrictions on switching and multi-homing of end users leads to a lack of contestability and reduced innovation.
\textit{Considerations:} 
The DMA prescribes portability of user data and 'continuous' and 'real-time' access to data~\cite[Art 6(9)]{dma}.
Data portability is further defined in the GDPR~\cite{gdpr}, where data is required to be provided in a structured, commonly used and machine-readable format. This was further clarified that data should include all data provided and generated by a user~\cite{DPWPguidelines}.

\subsubsection{\textbf{S6: Accessible Business Data}}
Business users are able to access their data and are enabled to request access to the data of their customers.
\textit{Problems:}
Business users and their end users generate vast amounts of data when using platforms. Gatekeepers are in a unique position to, e.g., withhold such data.
\textit{Considerations:}
The DMA prescribes to give business users `continuous' and `real-time' data access to their, and the directly related data of their users~\cite[Art 6(10)]{dma}. Where consent of the use of user data is required under the GDPR, the business user must be able to request such. 

\subsubsection{\textbf{S7: Accessible Training Data}}
Data collected on the platform, which is used to train algorithms, is made accessible to third-parties, without compromising user privacy.
\textit{Problems:} The vast amount of data collected by a platform constitutes an important barrier of entry and expansion for new competitors. This undermines the contestability of existing platforms.
\textit{Considerations:} The DMA requires search providers to enable third-parties access to `ranking, query, click and view data', which is generated through users. Access is required to be on `fair, reasonable and non-discriminatory terms' and personal data is required to be anonymised~\cite[Art 6(11)]{dma}.
%
\subsection{Transparent Practice}
Services offered on the platform are made transparent to allow involved parties to judge their fairness and weigh them against other offerings. This strategy focuses on the transparency of advertising services.
\subsubsection{\textbf{S8: Transparent Advertising}} Advertisement services on the platform are made transparent and verifiable in terms of performance and cost.
\textit{Problems:} Gatekeepers may exploit their position to provide non-transparent advertising services. The lack of information on e.g., fees, undermines the ability to make informed choices. 
Thus, the costs of services are likely higher due to decreased contestability.
\textit{Considerations:}
Advertisers and publishers are a type of business user, where the former places the advertisement (ad) in the `inventory' (e.g., published content) of the latter~\cite[Recital 45]{dma}.
The DMA requires both given access to tools and data to confirm the performance of an ad (including price and metrics for price calculation)~\cite[5(9-10)]{dma}. The DMA hereby requires enough data to enable 'independent verification'~\cite[Recital 45]{dma}.


%
%
\section{Discussion}
\label{sec:discussion}
\subsection{Strategies and Tactics}
We were able to derive tactics from the investigated gatekeeper compliance reports and associate them with our strategies derived from the DMA. 
In this short paper, we did not discuss the employed gatekeeper tactics, we plan to investigate them more closely in the future.


While implementation details may differ, gatekeepers align on the general tactics chosen to implement strategies. The noticeable exception is (S3) \textit{Equal Access \& Interoperability}, that pertains mostly to \textit{Alphabet} and \textit{Apple} (from our sample). Both have been known for different (or opposing) strategies on the degree of openness of their platforms~\cite{anvaari2010evaluating}. This also translates to their compliance approach. While Alphabet’s \textit{Google Android} operating system already exhibited a large degree of openness, Apple’s strategy to open their iOS operating system relies on (T6) \textit{Interoperability Requests}, which imposes requirements for services that seek deeper integration (as e.g., default apps). 
At the same time, (S3) \textit{Equal Access \& Interoperability} imposes the strongest demands on the technical design--specifically, if the system previously has not been designed around qualities of integrability or openness. This is testified by the many demands the DMA obliges on Apple's operating system \textit{iOS}\footnote{\label{fn:appleInter}See the specification decisions for \textit{iOS} and \textit{iPadOS}, \url{https://digital-markets-act.ec.europa.eu/questions-and-answers/interoperability_en}, accessed 2025-12-05} (which is known to follow a less open strategy~\cite{anvaari2010evaluating}).

We initially expected a dual-nature between the strategy group \textit{Equal Services} (S3 \& S4) and \textit{Transparent Practice} (S8). However, this has not materialised. For example, while the strategy (S4) \textit{Equal Ranking} also obliges a notion of transparency ('the conditions that apply to such ranking should also be generally fair and transparent'~\cite[Recital 52]{dma}), it seems to suffice that the ranking criteria is communicated. There is no technically enabled independent verification afforded (as it is, for example, demanded for ads). 


Throughout their compliance reports, gatekeepers voiced privacy, confidentiality, and security concerns as tradeoffs to the proposed tactics. Specifically, these were directly stated for tactics concerning interoperability (T5-T7, T9) and data access (T12–T15). 
The side-effects of our elicited tactics on quality attributes remains subject for future work. 

\subsection{Affordances}
While the DMA imposes obligations on gatekeepers, it also creates opportunities for new competitors entering the platform ecosystem. We, thus, extracted what affordances each tactic creates, i.e., which new design options emerge for third-parties based on a chosen tactic. An overview is given in Table~\ref{tab:tactics} (Affordance Perspective). These affordances can be divided in opportunities for increasing intra-platform competition (T4-T8, T10) and inter-platform (T9, T11-15) competition. 

\subsubsection{Intra-Platform Competition}
Affordances benefiting intra-platform competition create new design options for services, which so far were reserved for vertically integrated services. 
(T4) \textit{Default Service Management} enables apps or services to become the platform default for a user (e.g., browser, search engine). Tactics concerning interoperability (T5-T8) allow apps or services to better integrate into the platform and access new platform features. (e.g., hardware connectivity features previously restricted, as for example the interoperability features required to become available in Apple’s operating systems, see Footnote~\ref{fn:appleInter}.)
The implementation of (T10) \textit{Non-Preferential (Ranking) Criteria} will give equal exposure to services and apps, where previously platform services might have been preferred.

\subsubsection{Inter-Platform Competition}
Affordances benefiting inter-platform competition create opportunities for third-party platforms. (T9) \textit{E2EE Service Interoperability} will allow to integrate with an existing gatekeeper messenger service, which likely will decrease the cost of switching for users. Similarly, tactics enabling user data portability, access \& transfer (T11-T13) will enable the import of existing user data (e.g., moving profile and content to a new platform), reducing their lock-in on one platform. (T14) \textit{Provide Anonymised Training Dataset} enables access to training data, previously only available to gatekeepers with an established large user base. Finally, (T15) \textit{Two-Sided (Ad) Transparency}, enables the direct comparison of advertisement offerings in terms of cost, surcharges, and remunerations, promoting informed choice.

While, in principle, the elicited tactics offer these affordances, their effectiveness will depend on a gatekeepers implementation. 
%
\subsection{Values}
Based on the DMA’s recitals, we elicited a range of values associated with our strategies. Inherently, we cannot claim that any given strategy does not pertain exclusively to the attached value(s). Nor does the DMA offer concrete conceptualisations of these values that go beyond the definitions of contestability and fairness. A strategy will indirectly affect (promote or harm) other values. This will be the case for other values regarded in the DMA, but also other values not considered in the DMA.\footnote{The in the DMA expressed values are essentially contested concepts, and alternative conceptions exist. The DMA's perspective is referencing competition law.}
For example, as the DMA, in many cases, demands more openness, which is in natural opposition to values concerning privacy or confidentiality. The DMA specifically stresses that its implementation must be without prejudice to previous legislation like the GDPR~\cite[Recital 65]{dma}. Thus, it is imperative to capture such decisions early in design, so they can be aligned. 

Naturally, the DMA is itself an artefact with embedded values, which was formulated (designed) by (and under the influence of) different stakeholders and their interests. Consequently, one cannot reason about the DMA’s goals without considering geopolitical interests. 
There is an ongoing debate about the DMA’s enforcability and long term effects and whether the DMA's obligations can achieve the targeted goals~\cite{simone2025principles,bourreau2025interoperability,bostoen2023understanding}. Such a discussion, however, needs to take into account the technical operationalisation of these values.
Our strategies and tactics can be a starting point to also critically reason about this aspect by considering their influence for the high-level software architecture of a platform. However, future research is needed.
%
%
%
\subsection{Threats to Validity}
\label{sec:ttv}
\noindent We classify threats to validity according to Wohlin \etal \cite{WohlinEtAl_ExperimentationSoftware_2012}.

\textit{Construct validity.}
This study was executed by authors with training in software engineering and information systems; our interpretation of legal regulation may lack the nuance that legal researchers or practitioners would apply. Our strategies are not meant to achieve compliance (as it is always context dependent), but to provide high-level technical design strategies that can guide the architecture of such systems (c.f. the argument of Koops et al.~\cite{koops2014privacy}). To mitigate misinterpretation, we coded recitals, which already provide interpretation of the legally binding obligations. The obligations were used to cross-validate the completeness of our strategies. Further, we investigated compliance reports, which comprise the response of the actual target audience. The robustness of our conclusions remains limited with regards to triangulation with complementary data sources. However, as this would require `transdisciplinary research' it would go beyond the intended scope of this study.


\textit{Internal validity:} 
Standard threats of qualitative coding related to the subjective nature of results apply. While the coding, analysis, and matching in phases 1-3 were primarily conducted by the first author, the second author was consulted in cases of uncertainty. The second author then conducted an independent check to confirm or challenge the coding after finalisation. Compliance report coding in phase 4 was split among the first and second author. The other author was always consulted in cases of uncertainty.
Phases 4 and 5 were then executed by authors 1 and 2 together. Additionally, all results, i.e., strategies, values, gatekeeper tactics have always been jointly discussed.


\textit{External validity:}
Our study is inherently limited to the context of the DMA. As the DMA and compliance responses evolve, our results capture a snapshot of the current state of practice. However, we present a repeatable method so that our results can be updated, if required. 

\section{Conclusion}
\label{sec:conclusion}
We derived eight design strategies from the DMA that serve as fundamental approaches towards value-based architectural goals like 'fair practice', or 'user choice'. 
Additionally, we investigated public compliance reports from major platform businesses to derive tactics that show the current state of practice of designing for DMA compliance.
Our work is the first to holistically  investigate the technical implications of the DMA---one of the most ambitious recent EU legislations regulating digital transformation.
Our work is a starting point for guiding value-driven architecture design of new platforms, as well as for research on expanding the current state of the art in designing fair platforms. Future work must assess the use of such strategies in practice. 




\section*{Acknowledgment}
We wish to thank the Network Institute for providing an interdisciplinary forum in which we could discuss early results of this paper. Specifically, we want to thank Jaap Gordijn for his input.

\bibliographystyle{IEEEtran}
\bibliography{03_references}

\end{document}